\documentclass[11pt,showpacs,preprintnumbers,amsmath,amssymb,prd,nofootinbib,superscriptaddress]{revtex4-2}

\usepackage{dcolumn}% Align table columns on decimal point
\usepackage{bm}% bold math
\usepackage{ifpdf}
\usepackage{hyperref}
\usepackage{xcolor,color,graphicx,graphics,physics}
\usepackage[spanish,english]{babel}%, portuguese
\usepackage[latin1]{inputenc}
\usepackage[OT1]{fontenc}
\usepackage{latexsym,amssymb,amsmath,amsfonts, slashed,cancel, simpler-wick}
\usepackage{makeidx}
\usepackage{epsfig,subfigure}
\usepackage{natbib}
\usepackage{epstopdf}
\usepackage{mathrsfs}
\usepackage{hyperref}%\usepackage[colorlinks=true,linkcolor=blue]{hyperref}
\hypersetup{colorlinks=true, linkcolor=blue, citecolor=blue, urlcolor=blue}
\usepackage{enumerate}
\usepackage{tikz}
\usepackage{feynmp-auto}  % Alternativa para diagramas
\usepackage{tikz-feynman}
\tikzfeynmanset{compat=1.1.0}
\usepackage{fixmath}

\everymath{\displaystyle}
\usepackage{graphicx}

\usepackage[T1]{fontenc}
\usepackage{amsmath}
\usepackage{amssymb}
\usepackage{graphicx}
\usepackage{xcolor}

\newcommand{\bea}{\begin{eqnarray}}
\newcommand{\eea}{\end{eqnarray}}

\newcommand{\orcid}[1]{\href{https://orcid.org/#1}{\includegraphics[width=10pt]{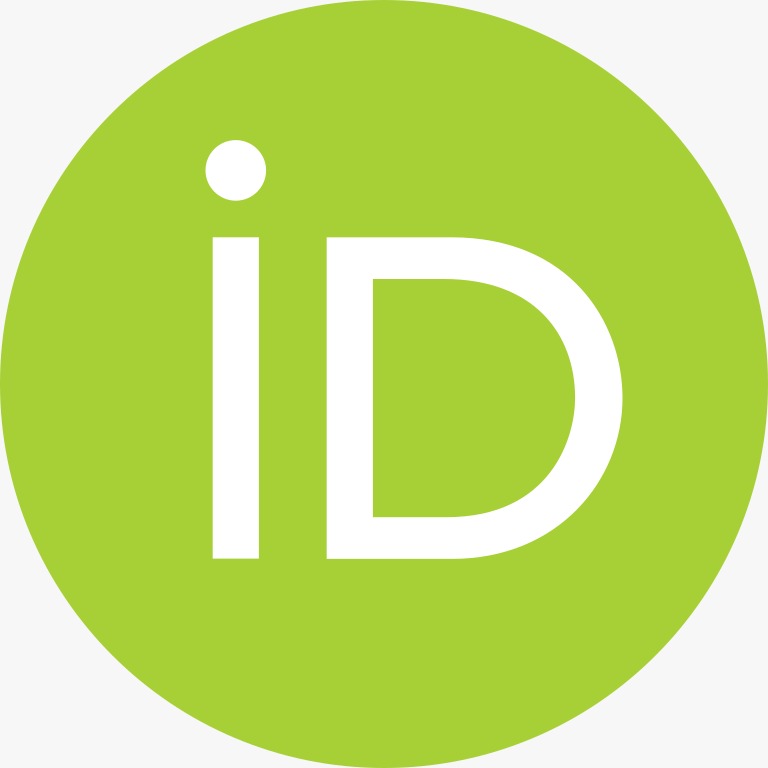}}}

\begin{document}

%\title{G\"{o}del and G\"{o}del-type metric in modified gravity}
%\title{G\"{o}del-type spacetimes and causality in $f(R,\mathcal{L}_m,\phi,X)$ gravity}
\title{Causality and its violation in $f(R,\mathcal{L}_m,\phi,g^{\mu\nu}\nabla_\mu \phi \nabla_\nu \phi)$ gravity}

\author{L. A. S. Evangelista \orcid{0009-0002-3136-2234}}
\email{lucassouza@fisica.ufmt.br}
\affiliation{Programa de P\'{o}s-Gradua\c{c}\~{a}o em F\'{\i}sica, Instituto de F\'{\i}sica,\\ 
Universidade Federal de Mato Grosso, Cuiab\'{a}, Brasil}

\author{M. L. R. Silva \orcid{0009-0002-0813-2502}}
\email{ludymilla@fisica.ufmt.br}
\affiliation{Programa de P\'{o}s-Gradua\c{c}\~{a}o em F\'{\i}sica, Instituto de F\'{\i}sica,\\ 
Universidade Federal de Mato Grosso, Cuiab\'{a}, Brasil}

\author{J. V. Moretti \orcid{0009-0000-2591-0455}}
\email{jmoretti@fisica.ufmt.br}
\affiliation{Programa de P\'{o}s-Gradua\c{c}\~{a}o em F\'{\i}sica, Instituto de F\'{\i}sica,\\ 
Universidade Federal de Mato Grosso, Cuiab\'{a}, Brasil}

\author{A. F. Santos \orcid{0000-0002-2505-5273}}
\email{alesandroferreira@fisica.ufmt.br}
\affiliation{Programa de P\'{o}s-Gradua\c{c}\~{a}o em F\'{\i}sica, Instituto de F\'{\i}sica,\\ 
Universidade Federal de Mato Grosso, Cuiab\'{a}, Brasil}

\begin{abstract}

A modified gravitational model whose action is given by an arbitrary function of the Ricci scalar, the matter Lagrangian density, a scalar field, and its kinetic term is investigated as an extension of the gravitational sector including an additional dynamical degree of freedom. Within this framework, the causal structure of rotating cosmological solutions is analyzed by considering G\"{o}del and G\"{o}del-type spacetimes as background geometries used as theoretical probes of the model consistency. Different matter sources are examined, including a perfect fluid and scalar-field configurations. It is found that the standard G\"{o}del metric is not compatible with the scalar sector of the theory unless the model reduces to the General Relativity limit. In contrast, G\"{o}del-type geometries admit a wider class of solutions whose causal properties depend on the model parameters and on the matter content. In particular, perfect-fluid sources may lead to either causal or noncausal configurations, whereas scalar-field configurations constrain the geometry to the causal limit, preventing the formation of closed timelike curves, highlighting that the scalar field plays a nontrivial dynamical role distinct from that of a cosmological constant.

\end{abstract}

\maketitle

\section{Introduction}

General Relativity is a well-established theory of gravitation and remains the most successful framework for describing gravitational phenomena. Over the past century, it has produced several remarkable predictions that have been confirmed by observations and experiments. Among its most notable achievements are the theoretical prediction and subsequent detection of gravitational waves, as well as the observational evidence supporting the existence of black holes \cite{einstein1916naherungsweise, einstein1918gravitationswellen}. At cosmological scales, these successes are naturally incorporated within the standard cosmological model, known as $\Lambda$CDM. Within this framework, a wide range of observations, such as the expansion of the Universe, the presence of dark matter, and the phenomena mentioned above, can be consistently described. Nevertheless, despite its impressive agreement with observations, this framework still faces important challenges, particularly in relation to the quantum description of gravity and the fundamental origin of the accelerated expansion of the Universe. In an attempt to address these issues, several alternative approaches have been proposed (see, for instance, the review in Ref.~\cite{lima2004alternative, clifton2012modified}), among which modified theories of gravity play a prominent role.

In general, such theories are constructed by extending or generalizing terms already present in the gravitational action. A well-known example is provided by $f(R)$ gravity models \cite{defelice2010fR}, in which the Ricci scalar in the Einstein--Hilbert action is replaced by an arbitrary function of the curvature scalar. In other approaches, the action functional is extended by introducing additional geometrical or matter contributions, or by establishing new couplings that are not present in the standard theory, as occurs in models such as $f(R,Q)$ gravity \cite{FRQ} and $f(R,T)$ gravity \cite{FRT}, among others. In recent years, modified gravitational theories have been extensively investigated in different cosmological and astrophysical contexts. In the present work, we consider a modified gravitational framework in which the action functional includes couplings between curvature, matter, and a scalar field treated as an additional dynamical degree of freedom in the gravitational sector. In particular, we adopt a general action whose functional dependence is given by $f(R,\mathcal{L}_m, \phi, g^{\mu\nu}\nabla_\mu\phi\nabla_\nu\phi)$ \cite{harko2018extensions, Harko_paper, Harko_2}. This construction makes it possible to investigate interactions between geometry, matter, and scalar fields within a unified framework, without assuming a specific cosmological origin for the scalar field, as well as to analyze their implications for the structure of G\"{o}del and G\"{o}del-type universes, which are employed here as theoretical backgrounds to probe the consistency of the modified field equations.

The G\"{o}del universe is a four-dimensional exact solution of the Einstein field equations that describes a homogeneous spacetime filled with pressureless matter and a negative cosmological constant \cite{godel1949example}. This solution is geodesically complete and possesses nonvanishing vorticity, which characterizes a global rotation of the matter distribution. As a consequence of this rotational property, the spacetime admits closed timelike curves (CTCs), allowing the possibility of causality violations. In this geometry, the light cones tilt due to the rotation, making it possible for worldlines to return to their own past. The G\"{o}del metric has been widely investigated in the context of modified gravity theories, including models such as $f(Q,T)$ \cite{ghorui2025godel}, $f(R,T)$ \cite{santos2013godel}, $f(R,Q)$ \cite{gama2017godel}, among others.

In the present work, the G\"{o}del metric is used as a test background to investigate whether the causality violations known to occur in General Relativity persist within the modified theory of gravity considered here. In addition, the analysis is extended to G\"{o}del-type geometries \cite{rebouccas1983homogeneity, rebouccas1985class, teixeira1985isometries, rebouccas1986note, calvao1988notes, santos2010godel}, which form a broader class of homogeneous rotating solutions characterized by two independent parameters, $m$ and $\omega$, that determine the causal structure of the spacetime. These geometries generalize the original G\"{o}del solution and allow the exploration of different regimes in which closed timelike curves may or may not arise depending on the relation between these parameters. In particular, the appearance of closed timelike curves is associated with the existence of a critical radius beyond which causality violations occur. By varying the parameters of the solution, G\"{o}del-type spacetimes provide a useful framework for investigating how modifications in the gravitational dynamics affect the causal structure of rotating cosmological models. G\"{o}del-type metrics have been studied in several contexts, for instance in Refs.~\cite{canuto2023godel, gonccalves2022godel, jesus2020godel, santos2015godel, agudelo2016godel}. Within this framework, we analyze the field equations of the $f(R,\mathcal{L}_m,\phi,g^{\mu\nu}\nabla_\mu\phi\nabla_\nu\phi)$ theory considering both a perfect fluid and a scalar field as possible matter sources, without addressing homogeneous and isotropic cosmological solutions, as our focus is on the consistency of G\"odel-type geometries rather than on cosmological evolution.

This paper is organized as follows. In Sec.~\ref{II}, the structure of the modified gravitational functional $f(R,\mathcal{L}_m,\phi,g^{\mu\nu}\nabla_\mu \phi \nabla_\nu \phi)$ is introduced and the corresponding field equations are derived. In Sec.~\ref{III}, the main properties of the G\"{o}del metric are reviewed and the modified gravity framework is applied to this background. In Sec.~\ref{IV}, a similar analysis is carried out for the G\"{o}del-type universe, and the resulting solutions are discussed with respect to the presence or absence of causality violations. In both analyses, different matter contents are considered. Finally, the conclusions are presented in Sec.~\ref{V}.

\section{Field equations of $f(R,\mathcal{L}_m,\phi,X)$ gravity}\label{II}

In this section, the field equations describing $f(R,\mathcal{L}_m,\phi,X)$ gravity are briefly introduced. This gravitational theory can be regarded as a generalized $f(R,\mathcal{L}_m)$ model. In this framework, the matter content of the Universe is composed not only of ordinary matter but also of a scalar field.

The action describing this model is given by
\begin{eqnarray}
 S=\int d^4x\,\sqrt{-g}\left\{ f(R,\mathcal{L}_m,\phi,X)+2\Lambda \right\},
\end{eqnarray}
where $g$ denotes the determinant of the metric tensor $g_{\mu\nu}$, $R$ is the Ricci scalar, $\mathcal{L}_m$ represents the matter Lagrangian density, and $\Lambda$ is the cosmological constant. The scalar field is denoted by $\phi$, and its kinetic term is defined as $X\equiv g^{\mu\nu}\nabla_\mu\phi\nabla_\nu\phi$. We note that the scalar field and its kinetic term could, in principle, be introduced as separate contributions to the action, rather than being incorporated into the functional dependence of $f$. Such a choice, however, would correspond to a different class of models with distinct dynamical properties. Furthermore, in order to ensure mathematical consistency, the function $f(R,\mathcal{L}_m,\phi,X)$ is assumed to be analytic in all of its arguments, so that it admits a Taylor expansion around a chosen background configuration.

By varying the action with respect to the metric tensor, we obtain
\begin{eqnarray}
\delta_g S=\int d^4x\left\{ f(\delta_g\sqrt{-g})+\sqrt{-g}\,(\delta_g f)+2\Lambda\,(\delta_g\sqrt{-g}) \right\}.
\end{eqnarray}
Here it has been assumed that $f=f(R,\mathcal{L}_m,\phi,X)$. The variation of the determinant of the metric tensor is given by
\begin{align}
	\delta_g \sqrt{-g}=-\frac{1}{2}\sqrt{-g}\,g_{\mu\nu}\delta_g g^{\mu\nu},
\end{align}
while the variation of $f$ can be expressed in terms of the variations of its arguments as
\begin{align}
	\delta_g f
	=f_R\,\delta_g R+f_{\mathcal{L}_m}\,\delta_g \mathcal{L}_m+f_\phi\,\delta_g \phi+f_{(\nabla\phi)^2}\,\delta_g (\nabla\phi)^2,
	\label{fall}
\end{align}
where, for convenience, we have defined $(\nabla\phi)^2=g^{\mu\nu}\nabla_\mu\phi\nabla_\nu\phi$ and $f_Y=\partial f/\partial Y$, with $Y=R, \mathcal{L}_m, \phi, (\nabla\phi)^2$. 

The variation of the Ricci scalar is given by
\begin{align}
	\delta_g R=R_{\mu\nu}\delta_g g^{\mu\nu}+\nabla_\mu\left(\nabla_\nu \delta_g g^{\mu\nu}-g_{\alpha\beta}\nabla^\mu \delta_g g^{\alpha\beta}\right).
\end{align}
The functional variation of $\mathcal{L}_m$ with respect to the metric tensor leads to
\begin{align}
	\delta_g \mathcal{L}_m=\frac{1}{2}\left(g_{\mu\nu}\mathcal{L}_m-\tau_{\mu\nu}\right)\delta_g g^{\mu\nu},
\end{align}
where $\tau_{\mu\nu}$ denotes the energy--momentum tensor. Since the scalar field is treated as an independent dynamical variable, its variation with respect to the metric vanishes, i.e., $\delta_g \phi=0$. Using $(\nabla\phi)^2=g^{\mu\nu}\nabla_\mu\phi\nabla_\nu\phi$, the variation with respect to the metric tensor yields
\begin{align}
	\delta_g (\nabla\phi)^2
	=\nabla_\mu \phi \nabla_\nu \phi\, \delta_g g^{\mu\nu}.
\end{align}
Collecting all contributions and imposing $\delta_g S=0$ for arbitrary $\delta_g g^{\mu\nu}$, the gravitational field equations read
\begin{align}
	f_R R_{\mu\nu}
	+\left(g_{\mu\nu}\Box-\nabla_\mu\nabla_\nu\right)f_R
	-\frac{1}{2}\left(f-f_{\mathcal{L}_m}\mathcal{L}_m\right)g_{\mu\nu}
	+f_{(\nabla\phi)^2}\nabla_\mu\phi\nabla_\nu\phi
	-\frac{1}{2}f_{\mathcal{L}_m}\tau_{\mu\nu}
	-\Lambda g_{\mu\nu}
	=0,
	\label{fieldeq}
\end{align}
where $\Box=g^{\alpha\beta}\nabla_\alpha\nabla_\beta$ is the covariant d'Alembertian operator.
These equations represent a generalization of Einstein gravity, allowing for nontrivial couplings between curvature, matter, and the scalar kinetic term.

To derive the scalar field equation, the action is varied with respect to $\phi$ while keeping the metric fixed. Then,
\begin{align}
	\delta_\phi S=\int d^4x\,\sqrt{-g}\,\delta_\phi f.
\end{align}
Since neither the Ricci scalar nor the matter Lagrangian density depends explicitly on $\phi$, only the explicit dependence on $\phi$ and the kinetic term contribute, yielding
\begin{align}
	\delta_\phi S
	=\int d^4x\,\sqrt{-g}\left[
	f_\phi\,\delta_\phi\phi
	+f_{(\nabla\phi)^2}\,\delta_\phi (\nabla\phi)^2
	\right].
\end{align}
Requiring $\delta_\phi S=0$ for arbitrary $\delta_\phi\phi$ yields the scalar field equation
\begin{align}
	\Box_{(\nabla\phi)^2}\phi
	=\frac{1}{2}f_\phi,\label{11}
\end{align}
where we have introduced the differential operator
\begin{align}
	\Box_{(\nabla\phi)^2}
	=\frac{1}{\sqrt{-g}}
	\partial_\mu
	\left(
	f_{(\nabla\phi)^2}\sqrt{-g}\,g^{\mu\nu}\partial_\nu
	\right).
\end{align}
This equation represents a generalized Klein--Gordon equation within this modified gravity framework.

In order to explore explicit solutions, the functional form of $f$ is specified as
\begin{equation}
	f = R + \mathcal{L}_m + \frac{\lambda}{2} g^{\alpha\beta} \nabla_\alpha \phi \nabla_\beta \phi,
\end{equation}
where $\lambda$ is a constant coupling parameter. In the present analysis, we restrict ourselves to the case of a vanishing scalar field potential. This choice is motivated by the need to simplify the field equations while isolating the effects associated with the kinetic term and the nonminimal couplings encoded in the functional dependence of $f(R,\mathcal{L}_m,\phi,X)$.

It is important to emphasize that the specific choice of the function $f$ adopted in this work corresponds to a restricted sector of the general $f(R,\mathcal{L}_m,\phi,X)$ framework. In this limit, the model can be interpreted as General Relativity supplemented by a scalar field with a nontrivial kinetic coupling. Although this reduces the overall level of generality, it enables a controlled analysis of the effects associated with the scalar degree of freedom.

We note that more general choices of the function $f$, including nonlinear dependencies on $R$, $\mathcal{L}_m$, or higher-order couplings involving $\phi$ and its derivatives, can also be considered. Such extensions would lead to a richer phenomenology and potentially to qualitatively different solutions. However, they would also significantly increase the complexity of the field equations, making an analytical treatment of G\"odel-type geometries considerably more challenging.
Consequently, the results obtained here should be understood as applying to this particular sector of the theory, which captures the essential features of the scalar--gravity coupling while remaining analytically tractable.

For this model, the derivatives of $f$ with respect to its arguments are given by
\begin{eqnarray}
\frac{\partial f}{\partial R} = 1, \quad
\frac{\partial f}{\partial \mathcal{L}_m} = 1, \quad
\frac{\partial f}{\partial \phi} = 0, \quad
\frac{\partial f}{\partial (\nabla\phi)^2} = \frac{\lambda}{2}.
\end{eqnarray}
Substituting these expressions into Eq.~\eqref{fieldeq}, the field equations reduce to
\begin{equation}
	2R_{\mu\nu}
	-(R+2\Lambda)g_{\mu\nu}
	+\lambda\nabla_\mu\phi\nabla_\nu\phi
	-\frac{\lambda}{2}g_{\mu\nu}g^{\alpha\beta}\nabla_\alpha\phi\nabla_\beta\phi
	-\tau_{\mu\nu}
	=0.
	\label{eq. de campo modificada}
\end{equation}

For this specific model, the scalar field equation \eqref{11} becomes
\begin{equation}
	\frac{1}{\sqrt{-g}}
	\partial_\mu
	\left(
	\frac{\lambda}{2}\sqrt{-g}\, g^{\mu\nu}\partial_\nu\phi
	\right)
	=0.
\end{equation}
Using the identities
\begin{align}
	\Gamma^\alpha_{\alpha\mu}=\frac{\partial_\mu\sqrt{-g}}{\sqrt{-g}},
	\qquad
	\partial_\mu g^{\mu\nu}
	=-\Gamma^{\mu}_{\mu\alpha}g^{\alpha\nu}
	-\Gamma^{\nu}_{\mu\alpha}g^{\mu\alpha},
\end{align}
the above equation can be written as
\begin{align}
	g^{\mu\nu}
	\left(
	\partial_\mu\partial_\nu\phi
	-\Gamma^\alpha_{\mu\nu}\partial_\alpha\phi
	\right)
	=0,
	\label{phiequation}
\end{align}
which corresponds to $\Box\phi=0$ up to the constant factor $\lambda$. Therefore, in this case the scalar field satisfies the standard covariant wave equation.

In the next section, the field equations will be applied to the G\"{o}del metric in order to investigate the implications of this model for rotating cosmological solutions.

\section{G\"odel metric}\label{III}

In this section, it is examined whether the G\"{o}del metric constitutes an exact solution of $f(R,\mathcal{L}_m,\phi,X)$ gravity. The G\"{o}del universe is a homogeneous and stationary solution of the Einstein field equations that describes a rotating spacetime filled with pressureless matter and a negative cosmological constant. Its nonvanishing vorticity generates a global rotation that tilts the light cones, allowing the existence of closed timelike curves (CTCs) and consequently leading to causality violation. This geometry therefore provides a useful framework for investigating the causal structure of the modified gravity model under consideration. The line element describing the G\"{o}del universe in Cartesian coordinates is given by
\begin{equation}
	ds^2 = a^2 \left( dt^2 - dx^2 + \frac{1}{2} e^{2x} dy^2 - dz^2 + 2 e^{x} dt dy \right),
\end{equation}
where $a$ is an arbitrary positive constant.

The nonvanishing Christoffel symbols for the G\"{o}del metric are given by
\begin{equation}
	\begin{aligned}
		& \Gamma^{0}_{01} = \Gamma^{0}_{10} = 1, \\
		& \Gamma^{0}_{12} = \Gamma^{0}_{21} = \Gamma^{1}_{02} = \Gamma^{1}_{20} = \frac{1}{2} e^{x}, \\
		& \Gamma^{1}_{22} = \frac{1}{2} e^{2x}, \\
		& \Gamma^{2}_{01} = \Gamma^{2}_{10} = - e^{-x}.
	\end{aligned}
\end{equation}
The nonvanishing components of the Ricci tensor are
\begin{equation}
	\begin{aligned}
		& R_{00} = 1, \\
		& R_{02} = R_{20} = e^{x}, \\
		& R_{22} = e^{2x}.
	\end{aligned}
\end{equation}
The Ricci scalar is
\begin{equation}
	R = \frac{1}{a^2},
\end{equation}
which is constant.

To solve the scalar field equation, the ansatz $\phi=\phi(z)$ is assumed. Under this assumption, Eq.~\eqref{phiequation} reduces to
\begin{align}
	g^{33}\phi^{\prime\prime}(z)-\Gamma^{3}_{33}\phi^\prime(z)=0.
\end{align}
Using the previous result, the solution of this equation is
\begin{equation}
\phi(z)=cz+d,
\end{equation}
where $c$ and $d$ are arbitrary constants. For this configuration, the kinetic term becomes
\begin{equation}
(\nabla \phi)^2 = g^{33} (\partial_3 \phi)^2 = g^{33} c^2 = -\frac{c^2}{a^2}.
\end{equation}

To solve the modified Einstein field equations (\ref{eq. de campo modificada}), the energy--momentum tensor must be specified. A perfect fluid is considered, whose energy--momentum tensor reads
\begin{equation}
	\tau_{\mu\nu} = (\rho + p) u_\mu u_\nu - p g_{\mu\nu},
\end{equation}
where $\rho$ and $p$ denote the energy density and pressure, respectively. The four--velocity is chosen as $u_\mu = (a,0,a e^{x_1},0)$, normalized with respect to the G\"{o}del metric.

For this matter content, and assuming that the scalar field depends only on $z$, the field equations reduce to
\begin{align}
	&1 - 2 a^2 \Lambda + \frac{\lambda c^2}{2} - \rho a^2 = 0, \\
	&1 + 2 a^2 \Lambda - \frac{\lambda c^2}{2} - p a^2 = 0, \label{system2_new} \\
	&3 - 2 a^2 \Lambda + \frac{\lambda c^2}{2} - 2 \rho a^2 - p a^2 = 0, \\
	&1 + 2 a^2 \Lambda + \frac{\lambda c^2}{2} - p a^2 = 0. \label{system4b_new}
\end{align}

By analyzing Eqs.~(\ref{system2_new}) and (\ref{system4b_new}), it is immediately found that the resulting system of equations is inconsistent. This inconsistency arises from the contribution of the scalar field. In particular, when $\lambda c^2 = 0$, the scalar-field kinetic term vanishes and the field equations reduce to those of general relativity. In this limit, the well-known G\"{o}del solution is recovered, for which the pressure and energy density are given by
\begin{align}
	p &= \frac{1}{a^2} + 2\Lambda, \\
	\rho &= \frac{1}{a^2} - 2\Lambda .
\end{align}

In order to investigate whether the system can be rendered consistent, alternative configurations may be considered. One possibility is to assume that the scalar field is the only matter source and depends exclusively on the time coordinate $t$. Another possibility is to allow the scalar field to depend simultaneously on temporal and spatial coordinates. In this latter case, the resulting field equations split into independent subsets associated with each type of dependence. However, each subset leads to consistency conditions analogous to those obtained for the case $\phi=\phi(z)$, ultimately leading to the same constraint on the metric parameters.

Therefore, within the model considered and for the class of scalar-field configurations discussed above, these results indicate that the G\"{o}del metric does not constitute an exact solution of the specific $f(R,\mathcal{L}_m,\phi,X)$ model considered here. It is evident that the exclusion of this solution is directly related to the presence of the scalar-field kinetic term. Consequently, in this context, this gravitational theory naturally prevents the existence of the G\"{o}del cosmological solution, which in general relativity is known to admit closed timelike curves and, therefore, violations of causality. In this sense, the scalar-field sector acts as a mechanism that effectively forbids G\"{o}del rotating solutions within the sector analyzed here.

In the next section, the same functional form of the theory is applied to a G\"{o}del-type metric. In contrast to the original G\"{o}del solution, the consistency conditions obtained previously are no longer required. Consequently, a different set of constraints on the model parameters and on the scalar-field configuration arises.

\section{G\"{o}del-type metrics}\label{IV}

To further investigate the occurrence of causal and noncausal regions in $f(R,\mathcal{L}_m,\phi,X)$ gravity, a generalization of the G\"{o}del universe, known as the G\"{o}del-type spacetime, is considered. As in the original G\"{o}del solution, this class of metrics represents a homogeneous rotating universe. However, unlike the G\"{o}del metric, G\"{o}del-type geometries are characterized by two independent parameters, $m$ and $\Omega$, whose values determine the causal structure of the spacetime and, in particular, whether closed timelike curves can occur.

The G\"{o}del-type spacetime metric, written in cylindrical coordinates $(t,r,\phi,z)$, is given by
\begin{equation}
	ds^2 = dt^2 + 2 H(r)\, d\phi \, dt - G(r)\, d\phi^2 - dr^2 - dz^2,
	\label{eq:godel_type_metric}
\end{equation}
where $G(r) = D^2(r) - H^2(r)$ and the functions $H(r)$ and $D(r)$ satisfy the spacetime homogeneity conditions
\begin{align}
	\frac{H'(r)}{D(r)}= 2\Omega, \quad \quad \quad \frac{D''(r)}{D(r)} = m^2,
	\label{hdequations_new}
\end{align}
with the prime denoting differentiation with respect to $r$. The parameters $\Omega$ and $m^2$ are real constants that characterize the G\"{o}del-type geometry, with $\Omega \neq 0$ and $-\infty < m^2 < \infty$.

The solutions of Eq.~\eqref{hdequations_new} define three distinct nondegenerate classes of G\"{o}del-type metrics, classified according to the sign of $m^2$, namely the hyperbolic, trigonometric, and linear classes \cite{rebouccas1983homogeneity}. In the present work, the analysis is restricted to the hyperbolic class, characterized by $m^2 > 0$ and $\Omega \neq 0$, which also includes the original G\"{o}del solution as a particular case. In this case, the metric functions take the form
\begin{align}
	H(r) &= \frac{4 \Omega}{m^2} \sinh^2\!\left(\frac{m r}{2}\right), \\
	D(r) &= \frac{1}{m} \sinh(m r).
\end{align}

The sign of $G(r)$ determines the causal structure of the spacetime. In particular, the existence of closed timelike curves is associated with the condition $G(r) < 0$. The critical radius $r_c$, beyond which these curves appear, is defined by
\begin{equation}
	\sinh^2\!\left( \frac{m r_c}{2} \right) > \left( \frac{4 \Omega^2}{m^2} - 1 \right)^{-1},
	\label{eq:relacao_rc_new}
\end{equation}
which constrains the parameters $(m,\Omega)$ and determines the region where causality is violated. It is important to emphasize that the particular case $m^2 = 2 \Omega^2$ reproduces the original G\"{o}del solution, for which the critical radius is finite and given by
\begin{equation}
	r_c = \frac{2}{m} \sinh^{-1}(1).
\end{equation}
In contrast, when $m^2 = 4 \Omega^2$, one finds $r_c \rightarrow \infty$, indicating the absence of closed timelike curves and, consequently, the preservation of causality.

To solve the field equations in the G\"{o}del-type spacetime, the Cartan formalism is adopted for simplicity. In a local frame, the metric (\ref{eq:godel_type_metric}) can be written as
\begin{equation}
	ds^2 = \eta_{AB}\,\theta^{A}\theta^{B}
	= (\theta^0)^2 - (\theta^1)^2 - (\theta^2)^2 - (\theta^3)^2,
	\label{eq:ds2_tetrad}
\end{equation}
where capital Latin letters are used to label local indices, $\eta_{AB} = \mathrm{diag}(1,-1,-1,-1)$ denotes the Minkowski metric, and the one-forms $\theta^{A} = e^{(A)}_{\ \ \alpha} dx^{\alpha}$ are defined by
\begin{align}
	\theta^0 &= dt + H(r)\, d\phi, \\
	\theta^1 &= dr, \\
	\theta^2 &= D(r)\, d\phi, \\
	\theta^3 &= dz .
\end{align}
Here $e^{(A)}_{\ \ \alpha}$ denotes the tetrad field, which satisfies the orthonormality condition	$e^{(A)}_{\ \ \alpha}\, e^{\alpha}_{\ (B)} = \delta^{A}_{\ B}.$ From the definition of the coframe, the nonvanishing tetrad components are
\begin{equation}
	e^{(0)}_{\ \ 0} = e^{(1)}_{\ \ 1} = e^{(3)}_{\ \ 3} = 1, 
	\qquad
	e^{(0)}_{\ \ 2} = H(r), 
	\qquad
	e^{(2)}_{\ \ 2} = D(r),
\end{equation}
which imply the following inverse components:
\begin{equation}
	e^{0}_{\ (0)} = e^{1}_{\ (1)} = e^{3}_{\ (3)} = 1,
	\qquad
	e^{0}_{\ (2)} = -\frac{H(r)}{D(r)},
	\qquad
	e^{2}_{\ (2)} = D^{-1}(r).
\end{equation}

In the local Lorentz frame, the nonvanishing components of the Ricci tensor take the form
\begin{align}
	R_{(0)(0)} &= 2 \Omega^2, \\
	R_{(1)(1)} &= R_{(2)(2)} = 2 \Omega^2 - m^2 .
\end{align}
The corresponding Ricci scalar is given by
\begin{align}
	R = 2 \left( m^2 - \Omega^2 \right).
	\label{eq:R_tetrad_new}
\end{align}
The nonvanishing components of the Einstein tensor in the orthonormal basis are therefore
\begin{align}
	G_{(0)(0)} &= 3 \Omega^2 - m^2, \\
	G_{(1)(1)} &= G_{(2)(2)} = \Omega^2, \\
	G_{(3)(3)} &= m^2 - \Omega^2 .
\end{align}

The final ingredient required to solve the field equations is the specification of the matter content. In the present analysis, the sources are taken to be a perfect fluid and a scalar field. We begin by considering the perfect fluid, whose energy-momentum tensor in the tetrad basis is given by
\begin{equation}
	\tau_{AB} = (\rho + p) u_{A}u_{B} - p \eta_{AB},
\end{equation}
where $u_A = (1,0,0,0)$ in the comoving frame.

With these definitions, the field equation \eqref{eq. de campo modificada} takes the form
\begin{align}
	G_{AB} = \frac{\lambda}{4} \eta_{AB} \eta^{MN} \nabla_{M} \phi \nabla_{N} \phi 
	- \frac{\lambda}{2} \nabla_{A} \phi \nabla_{B} \phi 
	+ \frac{\tau_{AB}}{2} + \Lambda \eta_{AB}.
	\label{field eq. tetrad f}
\end{align}

Using the results obtained previously, Eq.~\eqref{field eq. tetrad f} leads to the following set of equations:
\begin{align}
	3 \Omega^2 - m^2 & = - \frac{\lambda c^2}{4} + \frac{\rho}{2} + \Lambda, \label{00 f tetrad pf} \\
	\Omega^2 & = \frac{\lambda c^2}{4} + \frac{p}{2} - \Lambda, \label{11 f tetrad pf} \\
	m^2 - \Omega^2 & = - \frac{\lambda c^2}{4} + \frac{p}{2} - \Lambda. \label{33 f tetrad pf}
\end{align}

By subtracting Eq.~(\ref{33 f tetrad pf}) from Eq.~(\ref{11 f tetrad pf}), the following relation is obtained:
\begin{equation}
	m^2 - 2 \Omega^2 = - \frac{\lambda c^2}{2}.\label{godeltypesolution}
\end{equation}

It is important to note that Eqs.~\eqref{00 f tetrad pf}--\eqref{33 f tetrad pf} should not be interpreted as independent relations valid for arbitrary values of the energy density and pressure. Instead, they determine the matter content required to support the G\"odel-type geometry. In particular, the consistency condition given by Eq.~\eqref{godeltypesolution} depends only on the geometric parameters and is independent of $\rho$ and $p$. Once this condition is satisfied, the remaining equations fix the energy density and pressure.

The sign of the parameter $\lambda$ plays an important role in determining the causal structure of the G\"{o}del-type spacetime. From Eq.~\eqref{godeltypesolution}, it is found that for $\lambda < 0$ one obtains $m^2 > 2\Omega^2$. In this case, the existence of closed timelike curves depends on the relative magnitude between $m^2$ and $4\Omega^2$. As long as $m^2 < 4\Omega^2$, a finite critical radius $r_c$ exists and the spacetime admits noncausal regions. The limiting case $m^2 = 4\Omega^2$ corresponds to $r_c \rightarrow \infty$, yielding a completely causal configuration. This result is a distinctive feature of this gravitational theory, with no analogue in general relativity.

On the other hand, when $\lambda > 0$, the relation $m^2 < 2\Omega^2$ is obtained. In this regime, the inequality $m^2 < 4\Omega^2$ is automatically satisfied, which implies the existence of a finite critical radius and, consequently, the presence of closed timelike curves. Finally, in the limit $\lambda = 0$, the relation $m^2 = 2\Omega^2$ is recovered, corresponding to the standard G\"{o}del solution.

Let us now consider the case in which the matter content is entirely described by a scalar field, whose energy--momentum tensor is given by
\begin{align}
    \tau_{AB}= \partial_A \phi \partial_B \phi - \frac{1}{2}\,\eta_{AB}\,\eta^{CD}\partial_C \phi \partial_D \phi.
\end{align}
The scalar field is assumed in the form $\phi = \epsilon z + \epsilon_0$, where $\epsilon$ and $\epsilon_0$ are constants. Such configuration is adopted because it yields a constant energy-momentum tensor, consistent with the homogeneity of G\"odel-type spacetimes \cite{rebouccas1983homogeneity,rebouccas1985class}. More general configurations could be considered, but they would typically introduce coordinate-dependent contributions, potentially spoiling the symmetry of the solution and requiring additional constraints for consistency.

The nonvanishing components of the energy--momentum tensor associated with the scalar field are
\begin{equation}
    \tau_{(0)(0)} = -\tau_{(1)(1)} = -\tau_{(2)(2)} = \tau_{(3)(3)} = \frac{\epsilon^2}{2}.
\end{equation}

For this case, the field equations \eqref{field eq. tetrad f} reduce to the system
\begin{align}
	3 \Omega^2 - m^2 & = - \frac{\lambda c^2}{4} + \frac{\epsilon^2}{4} + \Lambda, \label{00 f tetrad s} \\
	\Omega^2 & = \frac{\lambda c^2}{4} - \frac{\epsilon^2}{4} - \Lambda, \label{11 f tetrad s} \\
	m^2 - \Omega^2 & = - \frac{\lambda c^2}{4} + \frac{\epsilon^2}{4} - \Lambda. \label{33 f tetrad s}
\end{align}

By combining Eqs.~(\ref{00 f tetrad s}) and (\ref{11 f tetrad s}), the following condition is obtained:
\begin{equation}
	m^2 = 4\Omega^2,
\end{equation}
which corresponds to the causal limit of the G\"{o}del-type class, since it implies $r_c \rightarrow \infty$. Therefore, when the spacetime is sourced exclusively by a scalar field, the G\"{o}del-type geometry is driven to the causal boundary of the solution space, preventing the formation of closed timelike curves within this gravitational framework. This result suggests that the scalar sector of the theory acts as a mechanism that suppresses the emergence of closed timelike curves in G\"{o}del-type geometries.

In summary, G\"{o}del-type solutions in $f(R,\mathcal{L}_m,\phi,X)$ gravity sourced by a perfect fluid can display either causal or noncausal configurations, depending on the value of $\lambda$ and on the relation between $m^2$ and $\Omega^2$. In contrast, when the source is purely scalar, the field equations enforce the condition $m^2 = 4\Omega^2$, placing the geometry at the causal boundary of the G\"{o}del-type class and excluding the existence of noncausal regions.

\section{Conclusions}\label{V}

The G\"{o}del and G\"{o}del-type metrics in the context of $f(R,\mathcal{L}_m,\phi,X)$ gravity have been investigated for a universe filled with a perfect fluid and a scalar field. In this framework, the gravitational functional depends on the Ricci scalar, the matter Lagrangian, a scalar field, and its kinetic term. This structure allows the field equations to be written in a form closely resembling those of General Relativity, with additional contributions arising from the scalar sector and from a modified gravitational coupling.

The scalar-field functional is chosen such that the corresponding scalar-field equation reduces to a Klein--Gordon-type equation. When this formalism is applied to the G\"{o}del metric, it is found that, independently of the matter configuration considered--whether a perfect fluid, a scalar field depending on the $z$ coordinate, or a purely temporal scalar-field configuration--the field equations consistently require the scalar coupling to vanish, $\lambda = 0$. Consequently, within this class of models, nontrivial scalar-field configurations are incompatible with G\"{o}del solutions.

However, this behavior does not persist when G\"{o}del-type geometries are considered. In this case, the modified field equations allow a broader class of solutions. For a perfect fluid source, the spacetime parameters satisfy the relation $m^2 - 2\omega^2 = -\frac{\lambda c^2}{2}$, indicating that the causal structure of the solution is controlled by the sign of the scalar coupling parameter $\lambda$. In particular, depending on the relation between $m^2$ and $\omega^2$, the spacetime may either admit or exclude the presence of closed timelike curves.

When the matter content is described solely by a scalar field, the system of equations leads to the condition $m^2 = 4\omega^2$. This relation corresponds to the limit in which the critical radius satisfies $r_c \to \infty$, implying the absence of closed timelike curves. Therefore, in this configuration the scalar field drives the solution to a completely causal G\"{o}del-type spacetime, suggesting that the scalar sector of the theory may act as a mechanism that suppresses the formation of closed timelike curves in this class of geometries.

%These results open several possible directions for future investigations. In particular, the present framework may be applied to other classes of rotating or anisotropic cosmological solutions in order to further explore the role of scalar fields in the causal structure of spacetime. Extensions of the model including more general scalar couplings, self interacting potentials, or different functional forms of $f(R,\mathcal{L}_m,\phi,X)$ may also reveal richer dynamical behaviors. Moreover, the connection between modified gravity effects and the suppression of closed timelike curves could provide further insights into mechanisms related to chronology protection in gravitational theories beyond General Relativity.

\section*{Acknowledgments}

This work by A. F. Santos is partially supported by National Council for Scientific and Technological
Development - CNPq project No. 312406/2023-1. L. A. S. Evangelista, M. L. R. Silva and J. V. Moretti thank CAPES for financial support.

\section*{Data Availability Statement}

%No data are available because of the nature of the research. This publication is theoretical work that does not require supporting research data.
No Data associated in the manuscript.

\section*{Conflicts of Interest}

No conflict of interests in this paper.

%%%%%%%%%%%%%%%%%%%%%%%%%%%%%%%%%%%%%%%%%%%%%%%%%%%%%%%%%%%%%%%%%%%%%%%%%%%%%%%%%%%%%%%%%%%%%%%%%%%%%%%%%%%%%%%%%

\global\long\def\link#1#2{\href{http://eudml.org/#1}{#2}}
 \global\long\def\doi#1#2{\href{http://dx.doi.org/#1}{#2}}
 \global\long\def\arXiv#1#2{\href{http://arxiv.org/abs/#1}{arXiv:#1 [#2]}}
 \global\long\def\arXivOld#1{\href{http://arxiv.org/abs/#1}{arXiv:#1}}

%%%%%%%%%%%%%%%%%%%%%%%%%%%%%%%%%%%%%%%%%%%%%%%%%%%%%%%%%%%%%%%%%%%%%%%%%%%%%%%%%%%%%%%%%%%%%%%%%%%%%%%%%%%

\end{document}